\documentstyle[12pt,aasms4]{article}

\def \ex{E^x}
\def \ey{E^y}
\def \ez{E^z}
\def \bx{B^x}
\def \by{B^y}
\def \bz{B^z}
\def \be{{\bf E}}
\def \bb{{\bf B}}
\def \et{{\bf \tilde E}}
\def \bt{{\bf \tilde B}}
\def \bv{{\bf v}}
\def \vt{v'}
\def \bvt{{\bf v'}}
\def \etx{{\tilde E^x}}
\def \ety{{\tilde E^y}}
\def \etz{{\tilde E^z}}
\def \btx{{\tilde B^x}}
\def \bty{{\tilde B^y}}
\def \btz{{\tilde B^z}}

\def \bj{{\bf j}}
\def \omxr{{\bf \Omega \times r}}
\def \gam{\gamma}
\def \gmunu{g_{\mu \nu}}
\def \gup{g^{\mu \nu}}
\def \Om{\Omega}

\def \csom{\cos \Omega t}
\def \csomp{\cos \Omega t'}
\def \snom{\sin \Omega t}
\def \snomp{\sin \Omega t'}
\def \xysq{x^2 + y^2}
\def \vr{\varrho}
\def \omxysq{\Omega^2 \vr^2}
\def \part{\partial}
\def \rl{r_L}

\def \qm{{q \over m}}
\def \mup{{\mu '}}
\def \nup{{\nu \,'}}
\def \xp{{x'}}
\def \yp{{y'}}
\def \zp{{z'}}
\def \gmunup{g_{\mup \nup}}
\def \alphap {{\alpha '}}
\def \betap {{\beta '}}
\def \sF{{}^* \! F}
\def \etal{{\it et al. }}

\begin{document}

\title{Electromagnetic Forces and Fields in a Rotating Reference Frame}
\author{Paul N. Arendt, Jr.}
\affil{Department of Physics, New Mexico Tech, Socorro, NM 87801}
\authoremail{parendt@aoc.nrao.edu}

\begin{abstract}
Maxwell's equations and the equations governing charged
particle dynamics are presented for a
rotating coordinate system with the global time coordinate of
an observer on the rotational axis.  Special care is taken in
defining the relevant entities in these equations.
Ambiguities in the definitions of the electromagnetic fields
are pointed out, and in fact are shown to be essential in such a system
of coordinates.
The Lorentz force is found to have an extra term
in this frame, which has its origins in relativistic mass.
A related term in the energy equation, which allows inertia
to be gained even during strict corotation, suggests ways
existing pulsar magnetosphere models may be modified to
match observed `braking indices' more closely.
\end{abstract}


\section{Introduction}

In pulsar magnetosphere theory and other applications \markcite{
ched96,chen84,fawl77,hon65,sch39} (Chedia
\etal 1996; Cheng 1984; Fawley, Arons, \& Scharlemann 1977;
Hones \& Bergeson 1965; Schiff 1939),
it is often convenient to adopt a rotating coordinate system
as a frame of reference.  Such a frame is non-inertial, and
has the usual Coriolis and centrifugal fictitious forces present
as a well-known consequence.  However, electrodynamic processes
are often of primary interest, and this leads to several difficulties.
In particular, the interpretation of various effects
and even the definitions of
the electric and magnetic fields are necessarily ambiguous
(some consequences of this were noticed by \markcite{back56}
Backus 1956).  It
is the purpose of this work to explore these matters in some
detail and thereby provide clarification.  In particular, we
will first present Maxwell's equations and then the Lorentz force
in a system of rotating Cartesian
spatial coordinates, with the global time coordinate of a (stationary)
observer on the axis of rotation.

As will be seen, it is difficult to define the electric and magnetic
fields in this frame.  This situation worsens with distance from the axis of
rotation, and becomes critical at the `light cylinder' distance,
$\rl = 1/\Om$, where $\Om$ is the angular frequency of rotation (here
and throughout, we use units where $c = 1$).  This situation will be
shown to be an essential feature of all such frames whose metric
tensor $\gmunu$ has off-diagonal elements.  For this reason,
all electromagnetic quantities used here will be given
careful definitions, with reference to their values in an inertial
non-rotating frame (where the ambiguities disappear).

Among the most surprising consequences which will be shown
are that the Lorentz force acquires an extra `relativistic mass'
term in this frame, and particles may gain inertia even when
undergoing strict corotation.
Although negligible at low altitudes,
these effects also can become critical at altitudes approaching
the light cylinder.  In pulsar models,
it is precisely in this regime where the magnetic
polar magnetospheric currents are expected to close; the details of
this process remain an open question whose answer is of key importance
for the models.  It is therefore hoped that this paper will
serve both as a reference for those who wish to adopt rotating
coordinates when studying electrodynamic processes, and as a guide
for interpreting the various phenomena in these coordinates.

\section{Definitions}

We must first define our coordinate systems; the remainder of
the mathematics is then straightforward tensor algebra.
To avoid an overabundance of `primes,' all coordinate-dependent quantities
with `primes' attached shall refer to the inertial (non-rotating)
frame, and `unprimed' quantities will be used for the rotating frame.
Greek letters will be used for indices which range over all
four spacetime components, for example, $U^\alpha = (U^t, U^x, U^y,
U^z),$ while
Latin letters will be used when variation is to be made over
the three ordinary spatial indices $(x,y,z)$.  The usual summation
convention is adopted.

We start with an inertial nonrotating Cartesian coordinate system
$(t',x',y',z')$ with the `flat' (ignoring gravity) Minkowski metric
$$ (ds')^2 = - (dt')^2 + (dx')^2 + (dy')^2 + (dz')^2 \eqno(1)$$
where $ds'$ is the infinitesimal proper distance for a spacelike
interval (and is coordinate-independent, although the `prime' is
left on for clarity).  If the interval is instead timelike, we
define its proper time  $d\tau'$ (also coordinate-independent) by
replacing $(ds')^2$ by $-(d\tau')^2$ above.  We
write the above as
$$(ds')^2 = g_{\mup \nup} dx^\mup dx^\nup, \eqno(2)$$
defining the metric tensor $g_{\mup \nup}.$

Next, we define coordinates which are rotating with respect to these, with
angular velocity $\Om$ counterclockwise about the $z'$ axis.  Thus,
$t = t',\> z = z',\> x = x' \csomp + y' \snomp,$ and $y = y' \csomp
- x' \snomp .$  Inverting gives $x' = x \csom - y \snom ,$ and
$y' = y \csom + x \snom .$  Note that the time coordinate is
identical to that in the nonrotating frame: we are
not measuring time as observers rotating with the frame would (which
we shall see has its advantages as well as disadvantages).  Naturally,
we will have an `ergosphere' beyond $r = \rl$, where all
material particles must have velocities opposing the rotation.

The differential coordinate transformation is
$dx^{\mu} = \Lambda^{\mu}_{\nup} \, dx^\nup ,$ which
defines the coefficients
$\Lambda^{\mu}_{\nup} \equiv \part x^{\mu} / \part x^{\nup}.$
The inverse transformation defines the coefficients
$\bar \Lambda_{\nu}^{\mup} \equiv \part x^{\mup} / \part x^{\nu},$
so that $\Lambda^{\mu}_{\nup} \, \bar \Lambda^{\nup}_{\sigma} =
\delta^{\mu}_{\sigma}.$

The metric $\gmunup$ transforms as a (symmetric) second-rank
covariant tensor:
$$g_{\mu \nu} = \bar \Lambda_\mu^\alphap \bar \Lambda_\nu^\betap
g_{\alphap \betap} \eqno(3)$$
which gives here
$$ \gmunu = \left( \matrix{-(1 -\omxysq) & -\Om \, y & \Om \,x &0 \cr
           - \Om \,y & 1 & 0 & 0 \cr \Om \, x & 0 & 1 & 0 \cr
           0 & 0 & 0 & 1} \right). \eqno(4)$$
(Although the use of cylindrical coordinates would obviously simplify
some expressions, such as this one for $\gmunu$, we stick to
rotating Cartesian spatial coordinates to avoid the added confusion
cylindrical coordinates give to covariant {\it vs.} contravariant
components of tensors.)
Notice that the coefficient of $dt^2$ vanishes on the cylinder $\omxysq
= 1 ,$ where $\vr = \sqrt{\xysq}$ is the distance from the rotation
axis (not to be confused with the charge density $\rho$).
However, the determinant of $\gmunu$ is always $-1$
due to the off-diagonal `space-time' terms, so there is
no mathematical difficulty in using these coordinates beyond
$\vr = \rl$.  This would not be true if we had picked the times
measured by local corotating observers as our time coordinate 
(as in \markcite{ched96} Chedia \etal 1996), which
would also have the undesirable property of having the definition
of $t$ vary with $\vr$.

We shall need the inverse of $\gmunu$.  It is
$$ \gup = \left( \matrix{-1 & -\Om \, y & \Om \, x & 0 \cr
- \Om \,y & 1- \Om^2 y^2 & \Om^2 xy & 0 \cr \Om \,x & \Om^2 xy & 1-\Om^2 x^2 &
0 \cr 0 & 0 & 0 & 1 }\right). \eqno(5)$$
The presence of nonzero off-diagonal terms in the
metric tensor and its inverse remind us that these coordinates are
not an orthogonal system, so that vectors and their
associated one-forms (covectors) do not generally `point' in the same
direction.   This will be shown to ultimately be
the reason why the notion of separate `electric' and `magnetic'
(spatial) vector fields runs into serious trouble in this frame.

\section{Maxwell's Equations}

The relativistically covariant way to discuss electrodynamics
in an arbitrary coordinate system is to introduce the antisymmetric
second-rank tensor $\bf F$ which has as covariant entries in the `primed'
(nonrotating) frame \markcite{mtw73,wein72,wald84}
(Misner, Thorne, \& Wheeler 1973; Weinberg 1972; Wald 1984)
$$ F_{\mup \nup} = \pmatrix{0 & -E^\xp & -E^\yp & -E^\zp \cr E^\xp & 0 &
B^\zp & -B^\yp \cr E^\yp & -B^\zp & 0 & B^\xp \cr E^\zp & B^\yp & -B^\xp &0}
\eqno(6)$$
where $E^{i'}$ are the (inertial-frame) components of the electric
field $\be$,
$B^{i'}$ are the components of the magnetic field $\bb$, and we
have assumed vacuum permeability and permittivity.
The components of $F_{\mu \nu}$ in any other frame are
then found in the same way as for $\gmunu$ in equation (3):
$$F_{\mu \nu} = \bar \Lambda_{\mu}^{\alphap} \bar \Lambda_{\nu}^{\betap}
F_{\alphap \betap} \eqno(7)$$
We find
$$F_{\mu \nu} = \pmatrix{
0 & -\etx & -\ety & -\etz \cr
\etx & 0 & \bz & - \by \cr
\ety & - \bz & 0 & \bx \cr
\etz & \by & -\bx & 0 }, \eqno(8)$$
where $E^i$ and $B^i$ are respectively defined as the 
projections of the electric and magnetic fields
(as measured in the inertial frame) onto the new (contravariant) spatial
coordinate vector basis.  For example, $\bb
= B^{i'} {\bf e}_{i'} = B^i {\bf e}_i$ (at corresponding
locations of the two coordinate systems; we are abusing notation),
where the ${\bf e}_i$ are the (spatial) contravariant coordinate basis for
vectors in the unprimed coordinates.  That is, ${\bf e}_i \equiv
(\part/ \part x^i),$ holding $x^j$ fixed for all $j \neq i.$  We
shall treat the $E^i$ and $B^i$ as numbers (rather than the
contravariant components of vectors) in what follows, for
clarity.   $\bf F$ is the fundamental tensor of interest, while
$\be$ and $\bb$ are not parts of 4-vectors and so do not
transform as such.

We have also defined the vector
$$\et \equiv \be + (\omxr) \times \bb \eqno(9)$$
which, from equation (8), seems to be the electric field in our
new coordinates.
However, we will also need the contravariant formulation of $\bf F$,
easily found by raising the indices on equation (8) with the
metric:
$$F^{\mu \nu} = \pmatrix{
0 & \ex & \ey & \ez \cr 
-\ex & 0 & \btz & - \bty \cr 
-\ey & - \btz & 0 & \btx \cr
-\ez & \bty & -\btx & 0 } \eqno(10)$$
where we have defined
$$\bt \equiv \bb - (\omxr) \times \be. \eqno(11)$$

It is straightforward (and reassuring) to verify that the invariants
$$\bb \cdot \bb - \be \cdot \be = {1 \over 2} F^{\mu \nu} F_{\mu \nu}$$
and
$$ \be \cdot \bb = {1 \over 4} F^{\mu \nu} \, \sF_{\mu \nu} $$
are reproduced by equations (8-11), where $\sF_{\mu \nu} \equiv
1/2 \, \epsilon_{\mu \nu \alpha \beta} F^{\alpha \beta}$ is the `dual'
tensor of $F,$ and $\epsilon_{\mu \nu \alpha \beta}$
is the totally antisymmetric
fourth-rank tensor (4-form) with $\epsilon_{0123} = 1.$

Notice that equations (9) and (11) do {\it not} represent a local Lorentz
transformation to our rotating coordinates, as our frame is
not locally Lorentz (due to the global time coordinate used).
Clearly, the interchanges
between the `normal' and `tilde' versions of $\be$ and $\bb$
when passing from the covariant to the contravariant representations
of $\bf F$ show a problem when trying to define the fields in these
coordinates.  We will see from Maxwell's equations
that we have no obvious way of choosing $\be$, $\et$,
or some combination of these objects as the `electric field' in this
coordinate system (for reasons to be detailed in section 5).
Some authors \markcite{chen84,fawl77,hon65}
(Cheng 1984; Fawley \etal 1977; Hones \& Bergeson 1965) 
simply follow the precedent set by \markcite{sch39}
Schiff (1939), who was resolving an interesting
paradox involving rotating frames and Mach's principle,
and who arbitrarily used the covariant
choices $\et$ and $\bb$.  Schiff correctly pointed out that it is the
components of the {\it mixed} tensor $F^\mu{}_\nu$
which dictate particle motion;
we shall return to this point after deriving Maxwell's equations for
the rotating frame.

Maxwell's equations in relativistically covariant form can
be written \markcite{wein72}
(Weinberg 1972)
$${\part (\sqrt{- g} F^{\alpha \beta}) \over \part x^{\alpha}} =
4 \pi \sqrt{- g} \, J^{\beta} \eqno(12a)$$
and
$$\epsilon^{\alpha \beta \gamma \delta} {\part F_{\gamma \delta}
\over \part x^{\beta}} = 0, \eqno(12b)$$
where $g$ is the determinant of $\gmunu$ (simply $-1$ here),
and $J$ is the current density 4-vector.  In the inertial frame, we have
${J'}^\alpha = ({\rho}',{\bj'})$ which transforms to
$J^\alpha = (\rho, \bj)$, where $\rho = {\rho}',$ and
$\bj \equiv \bj' - \rho(\omxr)$.   Note that the convection
of charge due to the rotation alters the effective current density.

Equation (12a) produces
$$\nabla \cdot \be = 4 \pi \rho \eqno(13a)$$
and
$$\nabla \times \bt = 4 \pi \, \bj + {\part \be \over \part t}, \eqno(13b)$$
while equation (12b) gives
$$ \nabla \cdot \bb = 0 \eqno(13c)$$
and
$$ \nabla \times \et + {\part \bb \over \part t} = 0. \eqno(13d)$$
Note the appearance of both `normal' and `tilde' components
of $\be$ and $\bb$.  Equations (13) constitute Maxwell's equations
in this frame.  Here, $\nabla$ is the usual 3-dimensional 
gradient operator, if rotating Cartesian coordinates are used.

We combine equations (13a) and (13b) to
get an equation for charge conservation:
$$ {\part \rho \over \part t} + \nabla \cdot \bj = 0 \eqno(14)$$
Equations (13) and (14) suggest that there is
no difficulty in interpreting $\rho$ and $\bj$ as charge and current
densities in this frame.  However, $J^\alpha$ is (unlike $\be$ and
$\bb$) a genuine 4-vector, with covariant components $J_\alpha
= \bigl( - \rho (1 - \omxysq) + (\omxr) \cdot \bj, \, \bj' \bigr).$

\section{Geodesics and Lorentz Force}

To explore particle dynamics, we use the geodesic equation, modified
by the Lorentz force when necessary.
First, we find the Christoffel symbols from the metric:
$$ \Gamma^\mu_{\alpha \beta} = {1 \over 2} \gup \bigg( {\part g_{\nu \alpha}
\over \part x^\beta} + {\part g_{\nu \beta} \over \part x^\alpha} - {\part
g_{\alpha \beta} \over \part x^\nu} \bigg). \eqno(15)$$
The only nonvanishing ones are found to be
$\Gamma^x_{tt} = -\Om^2 x,\, \Gamma^y_{tt} = -\Om^2 y, \,\Gamma^x_{yt} =
\Gamma^x_{ty} = - \Om,$ and $\Gamma^y_{xt} = \Gamma^y_{tx} = \Om .$

We now define the 4-velocity of a massive particle to be
$$U^\mu = {dx^\mu \over d\tau} \eqno(16)$$
where $\tau$ is the proper time of the particle.  Notice that this
gives $U^\mu U_\mu = \gmunu U^\mu U^\nu = -1 ,$ by the definition
of proper time.
A particle subject to no external forces travels along geodesics,
which are the solutions of
$$ {d U^\mu \over d\tau} + \Gamma^\mu_{\nu \lambda} U^\nu U^\lambda =
0. \eqno(17)$$

The temporal equation is trivial: $${d^2 t \over d\tau^2} = 0,$$ giving
$${dt \over d\tau} = {\rm constant} \equiv \gam.$$
We note that
$${d\over d\tau} = {dt \over d\tau} {d \over dt} =
\gam {d \over dt}, $$ so that 
$U^\mu = (\gam,\gam \bv),$ where $\bv = d{\bf x}/dt.$ Plugging this into 
$\gmunu U^\mu U^\nu = -1 $ gives, after rearranging,
$$\gam = {1 \over \sqrt{1- {\vt}^2}} \eqno(18)$$
where $\bvt \equiv \bv + \omxr$, the particle's velocity as measured
in the {\it nonrotating} frame of reference.  Thus, $\gam$ (which is
conserved in the absence of external forces) is just the particle
energy in the nonrotating frame (in units of $m c^2$).  Note, however,
that $\gam$ is not equal to $1/ \sqrt{1-v^2} ,$ which is fortunate
since $v>1$ is possible in this frame.

The spatial geodesic equations give (after cancelling a common $\gam$)
$$ {d \bv \over dt} =  - {\bf \Om \times (\omxr)} -2 ({\bf \Om \times v}).$$
This is just the familiar expression for the Coriolis and centrifugal
forces in a rotating frame. They are still valid relativistically,
even beyond the light cylinder.

When external forces are involved, we need only to introduce all
forces (per unit mass)
as 4-vectors $f^\alpha$,
and insert them on the right hand side of the geodesic equations in
a covariant manner: $dU^\alpha/d\tau + \Gamma^\alpha_{\mu \nu}
U^\mu U^\nu = f^\alpha .$

The Lorentz force is given by the contraction of $F^{\mu \nu}$ with
the 4-velocity $U^\alpha$:
$$ f^\alpha = \qm \gmunu F^{\alpha \mu} U^\nu \eqno(19a)$$
or equivalently by using the `mixed' representation of $\bf F$:

$$ {d U^\mu \over d\tau} + \Gamma^\mu_{\nu \lambda} U^\nu U^\lambda
= \qm F^{\mu}{}_\nu U^\nu, \eqno(19b)$$
where $q$ is the charge on the particle, and $m$ its mass.
The mixed tensor is here
$$F^\mu {}_\nu = \pmatrix{
\be \cdot (\omxr) & \ex & \ey & \ez \cr 
\etx - \ex \omxysq & \Om y \ex & \btz - \Om x \ex & - \bty \cr 
\ety - \ey \omxysq & - \bz - \Om x \ex & - \Om x \ey & \btx \cr
\etz - \ez \omxysq & \by & -\bx & 0 }.\eqno(20)$$

Substituting this Lorentz force into the geodesic equations gives
$$ {d \gam \over d t} = \qm \biggl[ \bv + (\omxr) \biggr] \cdot \be
 = \qm \bvt \cdot \be \eqno(21)$$
for the temporal equation, and
$$ {d (\gam \bv) \over dt} = \gam \biggl[ - {\bf \Om \times} ( \omxr )
- 2 \, {\bf \Om \times} \bv \biggr] + {q \over m} \biggl[ \et + \bv \times
\bb - (\bvt \cdot \be ) (\omxr ) \biggr]\eqno(22)$$
for the spatial equations.  Note the reappearance of $\gam$ in the
Coriolis and centrifugal forces, since $\gam$ is no longer conserved.

We call attention to the term
$$ (\omxr) \cdot \be $$
in equation (21).  It shows that a charged particle may gain
or lose inertia even when undergoing pure corotation, provided
there is a component of $\be$ in the direction of $\omxr$.  This
is obvious when considering particle motion in the nonrotating coordinates,
but is easy to forget when passing to the rotating frame, where the
`motion' becomes hidden.

We also see that the Lorentz force takes on a peculiar
form in our rotating frame.  The term $\et + \bv \times \bb$ seems to
lend credibility to naming $\et$ and $\bb$
as the electric and magnetic fields, but note that we could have also
have written it as $\be + \bvt \times \bb.$

What seems to be absent
in existing literature is the final term in the Lorentz force,
which can also be written as
$$ - (\omxr ) {d \gam \over dt}.$$
Note that the direction of this force is against (with) the rotation for a
particle which is gaining (losing) energy, betraying its origin
as an inertial `relativistic mass' effect.

\section{Interpretation}

We have already seen in equation (14) that $\rho$ and $\bj$ can be
naturally identified with the charge and current densities in
this frame; we wish to make similar identifications of the electric
and magnetic fields.  The standard technique from special relativity
determines the electric field by contracting $U^\alpha$ with the
mixed representation of $\bf F$:
$$ E^\alpha = (0,\be) \equiv F^\alpha{}_\beta U^\beta,$$
but this doesn't work here due to the nonzero component
$F^t{}_t$ in equation (20), which is also responsible for
the $(\omxr) \cdot \be$ term in equation (21).  (It is
interesting to note that
it is this `azimuthal' component of $\be$ for which
$\be$ and $\et$ always agree.)
Clearly, no prescription for `electric' and `magnetic'
fields will reproduce the Lorentz force as we are most
familiar with it.

The root of the problem is seen by considering equation (19b),
which says that the Lorentz force modifies geodesic motion by
`rotating' the tangent $U^\alpha$ to a particle's worldline
away from its parallel-transported value (along the worldline).
These `rotations,' by (the `unprimed' version of) the
metric equation (2), are restricted to satisfy the equation
$\gmunu U^\mu \, d U^\nu /dt = 0.$
In the flat Minkowski metric of special relativity, these rotations
are just the directions of Lorentz boosts and spatial rotations.  We
naturally identify forces which produce
spatial rotations of $U^\alpha$ as magnetic in nature, and
those which boost $U^\alpha$ as electric (for electrically charged
particles).  In this rotating reference frame, timelike and
spacelike components are not orthogonal to one another, so
boosts and rotations necessarily become mixed.  Indeed, 
infinitesimal Lorentz
boosts and rotations transform in exactly the
same way as do $\be$ and $\bb$, respectively.  We conclude from
this that
any reference frame with non-diagonal metric tensor $\gmunu$ will
suffer the loss of precise definitions of electric and/or magnetic
fields.

Nowhere is this more apparent than in equation (21), which allows
a charged particle's inertia to change provided $(\omxr) \cdot
\be$ is nonzero, even when the particle is at rest in the corotating frame.
Conservation of angular momentum demands that this exerts a torque on
the source of the fields ({\it e.g.}, the neutron star), which can alter
the frequency of rotation.  This phenomenon could be put to beneficial
use in pulsar models.  The `braking index' of pulsars, defined as
$$ n = {\Omega \ddot{\Omega} \over \dot{\Omega}^2} $$
(where the dots denote time derivatives)
has, when observable, always been found smaller than the 
canonically predicted
value $n = 3.$  (This is predicted as a {\it lower} limit when the
torque is from simple multipole radiation --- see, for example,
\markcite{kas94}
Kaspi \etal 1994 and references therein.)  Presently, the discrepancy
is understood as being due to particle outflow along lines of $\bb$, or
due to more complicated effects.  We see, however, that even the
corotating portion of a magnetosphere, with $\bb \cdot \be = 0,$
can alter the rotation rate by doing work on charged particles
(if $\be \cdot (\omxr) \neq 0$).

\acknowledgments

It is a pleasure to thank my colleagues Tim Hankins, Jean Eilek,
Jim Weatherall, David Moffett, and Tracey DeLaney for our weekly
pulsar discussions, which motivated the present work.  Funding
for this work was provided by NSF grant AST-9315285.

\end{document}